\documentclass[10pt]{article}

\usepackage[utf8]{inputenc}

\usepackage{amsmath}
\usepackage{cancel}
\usepackage{xcolor}
\usepackage{mdframed}
\usepackage{graphicx}
\usepackage{geometry}

\usepackage{sectsty}
\sectionfont{\fontsize{14}{25}\selectfont}

\title{Noisy Machines}

\author{
  Michael Kewming\\
  \\
Centre for Engineered Quantum Systems \\The School of Mathematics and Physics\\ The University of Queensland\\ QLD 4072 Australia
}

\date{}

\begin{document}
\maketitle

\begin{quote}
\emph{Mathematics is not a science from our point of view, in the sense that it is not a natural science. The test of its validity is not experiment.}
    \\
    \flushright{--- Richard Feynman \\ The Feynman Lectures on Physics}
\end{quote}

\subsection*{It's a Noisy World}

Complexity exists at every scale, whether we look upward at the skies or below the surface of our feet.
Whether it be understanding the genome, modern computers, or the weather, one thing is true: we live in highly uncertain, unpredictable and largely inscrutable universe.
Our predictions typically require simplification and averaging, allowing us to develop linear narratives which factorise neatly in our scientific understanding.
This is the picture that modern science has unearthed.

In complex systems errors compound and information rapidly dissipates through frequent interactions with the other subsystems.
At every step, randomness enters the picture creating a world that is fundamentally \emph{noisy}.
Even the deepest regions of space are never truly quiet.
The afterglow of the big bang buzzes in the background and virtual particles  pop in and out of existence. 
No matter where you are in the universe, you cannot escape noise. 
The laws of thermodynamics and quantum mechanics ensure dissipation and random fluctuations will follow your predictions everywhere.

One might seek peace and quiet in the abstract, and turn to pure mathematics and symbolic logic.
Bertrand Russell described mathematics as possessing ``not only truth, but supreme beauty---a beauty cold and austere, like that of a sculpture''.
Russell made these remarks in 1907 prior to the publication of Principia Mathematica.
Kurt G\"{o}del showed that mathematics as a closed axiomatic system can be neither complete and consistent---there must be at least one unprovable axiom, or a single contradiction: The marble edifice contained cracks.
Alan Turing traced these fissures through to their logical conclusion.
Turing envisioned mechanical devices---a Turing machine---capable of processing input signals, executing logical functions and then writing the results. 
In short, a Turing machine is an attempt to mechanise mathematics. 
Logically, he imagined a universal machine capable of emulating all of mathematics. 
Like G\"{o}del, Turing turned this system on itself and asked the seemingly simple question `is there a function it could not compute'? 
His answer was the baffling Halting problem: Assuming a given input and problem, could the Turing machine determine whether it would return a definite answer, or continue to run indefinitely? 
Turing showed this meta problem is \emph{undecidable}. 
Moreover, his work linked the unprovability of G\"{o}del's theorems to the new notion of undecidability.
Needless to say his work laid the foundations of modern computers, proving that these results do not compromise the utility or beauty of mathematics, but rather limit the capacity of mathematics to describe itself.

So where does this leave us?
Nature appears to be unpredictable and uncertain because of thermodynamics and quantum mechanics.
Mathematics, as a closed axiomatic system appears to be unprovable and undecidable.
These two perspectives live on opposite sides of a Platonic divide; the difference between objects (nature) and forms (abstractions). 
But Turing machines are mechanical by definition so a physical realisation of mathematics should indeed be possible. 
As Fredkin and Toffoli put it \cite{fredkin_conservative_1982}
\begin{quote}
    \emph{Computation--whether by man or by machine--is a physical activity, and is ultimately governed by physical principles.}
\end{quote}
This sentiment dates back to Szilard's exorcism of Maxwell's demon and the work of Landauer \cite{landauer_wanted_1967} who contrived the punchy line \emph{information is physical} \cite{landauer_physical_1996}.

This idea can be demonstrated rather simply;
let us try to construct an infinite loop like the Halting problem on a laptop.
Our CPU requires energy to perform the underlying calculation.
Thankfully we have a battery as an energy source.
As the CPU processes this calculation, it heats up due to the dissipative nature of several logical primitives.
Our laptop is currently erasing its memory to avoid clogging up with information, causing it to heat up further.
All this heating increases the error rate in the processing unit.
An underlying architecture of error-correction kicks into gear to preserve the fidelity of our computation.
This consumes further energy slowly draining our battery.
We plug the laptop into the wall where it draws energy from the local power station to continue on with the calculation.
Eventually enough time will pass where the computer components begin to degrade and fail.
Our only option is to replace these failing parts in real-time without interrupting the calculation.
And so on until the end of the universe.

What happens to your calculation if a single link in this complex chain breaks? 
The infinite loop is going to halt.
This example reveals a simple principle; closed physical systems only have access to finite pool of resources and will always require an external environment to source their energy and dissipate heat.
In the world of finite systems, nature will always force a solution to the Halting problem because nature does not permit contradictions: The asymptotic predictions of closed axiomatic systems can never exist in the physical world.
This pertinent observation can be gleaned through the historical conversations surrounding the construction of computers.
A computer is an attempt to build a closed logical system, embedded in the physical world. 
But just like a steam engine, it requires an energy source which necessarily slaves it to the universal laws of thermodynamics.
All physical computers, whether vacuum tubes or modern quantum computers require external power supplies. 
Coupling to an external environment ensures noise and dissipation are constants which must be handled by error-correction.

Halting problems and Universal Turing machines may exist in the realm of abstractions---that cold austere realm of vectors, perfection and infinity---but physically they are fantasies. If this abstract realm exists, we can never access it with physical Turing machines. All physical systems require energy which must be sourced from an environment and repaid with entropy; a universal truth upheld by the second law of thermodynamics.

\subsection*{Determinism vs In-determinism}
We should then carefully delineate between two broad classes of physical theories; deterministic and in-deterministic theories.

Deterministic theories, like those of classical mechanics and quantum mechanics, deal with the dynamical evolution of a completely specified microscopic state. 
In both theories of classical and quantum mechanics, the evolution of the underlying dynamics is predicted by a known differential equation i.e the Lagrangian, Hamilton or Schr\"{o}dinger equation. 
Moreover, both cases of deterministic theories are perfectly time \emph{reversible}.
If we start with a state $\psi(t_{0})$ and compute it at later time $\psi(t_{f})$, we can reverse the temporal direction and recover the initial state $\psi(t_{0})$.
From the perspective of thermodynamics, a reversible processes is one that remains in thermodynamic equilibrium with its environment. 
Consequently, no information is lost and every interaction is accounted for.
In either case there is one constant; entropy does not increase.
Reversible processes are of course an asymptotic approximation of reality, albeit a very useful one.
Such an approximation is only valid when the separation of time scales between the system and its environment are large. 
Deterministic theories like a microscope, focus down on a highly specific piece of nature by closing out the rest of reality.

In-deterministic theories on the other hand describe the macroscopic behaviour of an entire ensemble.
Statistical mechanics and kinetic theory are examples of such in-deterministic theories.
They provide powerful insights into the epiphenomena emerging from large numbers of interacting subsystems.
Curiously, emergent properties such as temperature, pressure and entropy have become subject to new deterministic models such as the ideal gas law and the laws of thermodynamics.
Most importantly, in-deterministic theories are \emph{irreversible} with regards to the underlying microstate.
Sometimes called out-of-equilibrium dynamics, irreversible processes arise when the separation of time scales between a system and its environment is small. 
Thus, interactions between the two cannot be discounted.
If the environment is not constantly monitored, these interactions will manifest as dissipation; the transfer of heat (energy) between the two.
Dissipation produces entropy at a finite rate and underpins the second law of thermodynamics. 
In complex systems with millions of interacting parts, it is computationally intractable to trace the dynamics of every degree of freedom.
We must coarse grain our observations, the price of which is a probabilistic description of the underlying microstates and the introduction of noise.

These two approaches are colloquially known as the theories of closed and open systems respectively.

\subsection*{Mechanical Mathematics}
Is mathematics physical? 
For many, the answer is a definite no and it is something more fundamental.
Mathematics is an idealised set of abstract forms divorced from physics.
Unsurprisingly many disagree with this dualism---most notably physicists.
Rather pragmatically, they posit that because physical systems perform the abstract task of mathematics---pens, pencils, and brains---then the emergent phenomenon of mathematics must too be physical.
However, this perspective holds for the \emph{application} of mathematics i.e bridging the Platonic divide and phyiscally realising mathematical abstractions. 

One of the most illuminating perspectives on the physicality of mathematics originates from the discussion surrounding the construction of real world Turing machines; that swinging bridge between the abstract and mechanical.

Rolf Landauer synthesised this attitude succinctly when he wrote ``that the tools of mathematics are physical, and therefore mathematics is limited by nature'' \cite{landauer_wanted_1967}.
This reasoning---now fairly widespread---is primarily tied to the development of modern computers. 
The logical processing of mathematics requires physical objects to carry out mathematical operations which are subject to physical constraints.
Landauer proposed the minimum energetic cost of irreversible computation in a system at temperature $T$ is $k_{B} T \ln 2$; this principle now bears his namesake \cite{landauer_irreversibility_1961}.
Others have since identified similar constraints pertaining to the processing rate of physical information \cite{bekenstein_energy_1981} through to the computational capacity of the universe \cite{lloyd_computational_2002}.

In Roger Penrose's best selling work \emph{The Emperor's new mind}, he wrote that a Turing machine is ``a piece of `abstract mathematics' and not a physical object" \cite{penrose_emperors_2002}. 
Externally, this idealised machine has the ability to draw from an unbounded pool of resources---or as he calls it `rough paper'.
Internally it contains a finite number of discrete states, but is capable of reading, writing and processing this infinite amount of rough paper.

Penrose confessed that this construction is a little uncomfortable since shifting an infinite piece of paper might be technically impossible. 
Instead, he circumvents this peculiarity by proposing that rough paper could be an external environment which the machine is embedded in. 
The machine reads and writes from the environment in a perfectly deterministic fashion.
This view provides a useful picture regarding the construction of a Turing machine in a physical world, but is equivalent to that old English adage of `having your cake and eating it too'. 
The Turing machine is treated as a deterministic closed system but sources its energy from an external environment. 
By definition, this is impossible to do without introducing errors.
Penrose's machine cannot physically do work on any input/output without an environment doing work on the system. 
This necessarily implies the Turing machine is not in thermal equilibrium with its environment.
Therefore, this theoretical description of a physical Turing machine is both dissapative and in-deterministic.
Any result relying on an unphysical assumption such as the deterministic sourcing of energy from an environment must be taken with a grain of salt or we must permit, by the same token, the construction of perpetual motion machines.

A more thorough picture espoused by Fredkin and Toffoli considers the irreversible nature of logical primitives required in modern \emph{classical} computing \cite{fredkin_conservative_1982}.
Several of these primitives---for example, the AND gate---are intrinsically irreversible.
In their words ``performing the AND operator one generally erases a certain amount of information about the system's past''.
This is the physical reason your CPU requires constant cooling.
All this `forgetting' generates heat which must be dumped back into the environment.
To avoid all this waste, they proposed a new microscopic and \emph{reversible} model of computation which they termed `conservative logic'. 
I will not elaborate on the technicalities of this article, but they assert that dissipationless computation is indeed possible and the energy involved is no greater than what is required to read/write the input/output signals.
The underlying process of abstract mathematics could be done---in theory---free of charge.
In this sense, the physical restrictions of doing mathematics was not in the mathematics itself, but rather encoding it in a physical structure.

Despite being free of energy consumption, conservative logic---like its dissipative counterpart---still requires a foundation of error-correction or post-selection to handle internal/external errors \cite{bennett_thermodynamics_1982}.
This landmark result along with the work of Benioff \cite{benioff_computer_1980} kicked off the quantum computing race and was an early link between quantum mechanics and Turing machines, a theme we will return to shortly. 

The limitations I discussed in Penrose's picture apply in part to Fredkin and Toffoli's result.
Their conservative logic proposition still requires energy consumption when embedding the logic in a physical structure.
As we will discuss, this process is irreversible, even down at the quantum mechanical limit.
While the underlying mechanics might indeed be reversible, coupling to an environment ensures noise will creep in and errors will sabotage the calculation.
If the external world is to be the source of energy, then doing the abstract task of mathematics still requires we model the \emph{whole} process as intrinsically irreversible.

\subsection*{Error-correction}

Let's assume that we can construct a Halting problem on a physical Turing machine $T_{H}$.
Its central processing unit is intrinsically reversible but is still embedded in an \emph{infinitely large} physical environment. 
As we discussed, energy is still required to read and encode the data.
Critically, this opens the processing unit up to noise which generate errors and corrupt your Halting calculation; for some unexplained reason, your Turing machine is going to randomly halt. 
A remedy would be to include a third system, another Turing machine $T_{E}$ embedded in the same environment which constantly performs error-correction on $T_{H}$. 
By doing so, $T_{H}$ can be certain that the data it is processing is error free.
Ideally the Turing machine $T_{E}$ never makes an error, but given it too is embedded in the same physical environment it will make errors from time to time.
Unsurprisingly, it too must be monitored by another system $T_{F}$.
Error-correction is thus, a necessary component of physical Turing machines.

Error-correction fundamentally relies on a computational space larger than the subspace performing the calculation---otherwise known as the \emph{code space} \cite{nielsen_quantum_2010}. 
The input signal is encoded into the Turing machine, which casts the logic into larger mathematical space. 
This new space must be large enough to distinguish between different types of errors.
For example, if our input tape was just a binary string of $0$s and $1$s, these would be called \emph{bit flip} errors where $0\rightarrow1$ and vice versa.
To detect simple bit flip errors, we require at least two additional bit per encoded bit for successful error-correction.

In principle, a Universal Turing machine should possess error-correcting logic in its repertoire of functions.
Thus we could draw a larger box around $T_{H}$ and $T_{E}$ and call this joint system our Universal Turing machine. 
An interesting digression is the following consideration; inside our Turing machine we have split the logic functionality in two; half of our Universal Turing machine is processing the self-referential logic whereas the other half is simply error-correcting. 
In this sense, self-reference in physical systems does not refer to the self as a whole, but rather only the subsection, or more precisely the code space.
But let's not be too hasty and forget about the errors arising in $T_{E}$ which might also make errors. 
Naively, we could just extend the Turing machine indefinitely to include more and more parts, each of which is monitoring another part of the Turing machine ad infinitum.
However, a more physically realistic argument would be that Turing machines can only support Halting-like problems in a physical world by relying on the external network of other machines providing error-correction.

But what game are we playing here? 
Like sailors on a sinking ship, we are frantically trying to plug leaking holes of noise. 
With the introduction of more and more error-correcting units, we are merely patching hole after hole but never truly closing the system.
Also, error-correction does not only apply to the software of the Turing machine. 
With the passing of time, the degradation of the physical Turing machine is going to introduce further errors.
Only constant repair and replacement of parts is going to keep this undecidable calculation running---another form of error-correction. 
Error-correction is thus an attempt to stave off the unalterable march of the second law of thermodynamics.

\subsection*{Quantum Reality}
Quantum physics is one of modern science's crown jewels---despite its peculiarities.
Any physical realisation of a Turning machine must comply with the rules of quantum system. 
We are thus lead to the Church-Turing-Deutsch (CTD) principle, which succinctly states \cite{deutsch_quantum_1985}:
\begin{quote}
     \emph{Every finitely realisable physical system can be perfectly simulated by a universal quantum computer operating by finite means.}
\end{quote}
Note that the CTD principle is not a thesis like the Church-Turing thesis.
The difference being the CTD principle is subject to constraints such as physical systems, measurements and encodings.
Quantum computers---like Turing machines---are capable of executing mathematical logic on quantum mechanical hardware.
Their processing logic is entirely reversible as it's evolution is governed by the Schr\"{o}dinger equation.
This hardware offers fundamental resources that are off-limits to classical computers, including entanglement, discord and superposition.
Furthermore, the inherent uncertainties arising from Heisenberg uncertainty principle need not put constraints on reversible logic gates \cite{milburn_quantum_1989}.
This is why quantum computers are the subject of the CTD principle and not classical computers.
In characteristic fashion, Feynman tersely summarised this idea in his infamous 1982 talk; ``nature isn't classical, dammit, and if you want to make a simulation of nature, you'd better make it quantum mechanical'' \cite{feynman_simulating_1982}.

In saying this, we should not delude ourselves that quantum computers are omnipotent computing machines.
While the underlying hardware is reversible and their computational ability far outstrips that of their classical counterparts, the full mechanical picture of quantum computation is still \emph{irreversible}.
Building a universal quantum computer capable of supporting a Universal Turing machine would be subject to the same limitations outlined above.
Dissipation enters a quantum computer through the same routes as a classical computer, via the process of encoding, measurement and error-correction.
When a quantum system is prepared, an external probe must instantiate its quantum state.
When a quantum system is measured, it's wave function collapses---in the most common interpretation---and randomly returns a value associated with the measured observable according to some probability distribution\footnote{I will not digress into a rather contentious discussion regarding the interpretations of quantum mechanics here, but will rely on the operational output of the theory.}.
These ideas are deeply rooted in language of quantum physics.
We talk about `opening' up Schr\"{o}dinger's box or `reading' off the state of the wave function.
The quantum computer must be opened up temporarily for preparation and observation via an experimental apparatus.

The work of Toffoli and Fredkin is foundational here. 
Operated at its most advantageous limit, a quantum computer still requires an energy source to encode and decode the inputs and outputs. 
The energy source must come from an external environment out of thermal equilibrium with the system and thus, subjects the quantum device to similar thermodynamic constraints as it's classical counterpart \cite{strasberg_quantum_2017}.
This environment plays the role of the irreversible ``observer'' in quantum mechanics: It also serves as the experimental apparatus.
Moreover the coupling between the system and the apparatus is \emph{finite} and limited by the Heisenberg uncertainty principle; their can be no instantaneous or perfectly localised measurements in quantum physics.
The price of this delineation is an increase in entropy and the introduction of noise.
Internally, we can rectify these errors by introducing further ancillary systems within our computer whose sole function is to perform error-correction. 
These ancillary systems are periodically measured and condition the wavefunction, ensuring it executes error free.
But again, we are playing this same game of trying to close a system from an external world that it draws its energy from.

Finally a remark regarding those who subscribe to ``the church of the higher Hilbert space'' i.e extending the boundaries of the environment to close out the system. 
The coupling between system and apparatus I have described draws a hard boundary delineating the two, otherwise known as the \emph{Heisenberg cut}.
Taking a microscopic picture, the system and the apparatus become entangled i.e the quantum mechanical wavefunction of the two cannot be separated. 
In this context the apparatus and the system form a new closed system in a superposition of all possible prepared and measured outcomes.
However, if the environment is large then the sensitive coherent interaction between these two is rapidly dissipated. 
There is a vanishing probability that the quantum correlations between the system and environment can be preserved in a thermal bath.
This is known as a the Markov assumption, whereby the environment is viewed as conditionally independent of the system.
Furthermore, recent research suggests that gravity introduces a further source of decoherence \cite{pikovski_universal_2015}, which is closely related to the gravitational collapse models espoused by Penrose \cite{penrose_gravitys_1996}.
Because gravity cannot be screened off like other interactions such as electromagnetism, it implies no sub-system can ever be truly isolated.
Even if you manage to close your system off from a noisy external environment, the canvas of spacetime may introduce it for you.

\subsection*{Conclusion}
All asymptotic predictions of closed axiomatic systems can never exist in the physical world.
The abstract notion of a Universal Turing machine cannot exist as a physical subsystem without the introduction of noise from an external energy source.
While it is theoretically possible to build a processing device which is fundamentally non-dissapative such as quantum computers, the process of inputting and outputting information requires energy and introduces noise. 
Additionally it requires an internal mechanism of error-correction to make the program robust against errors.
The processing logic of error-correction must come from an external processing unit also sourcing its power from the environment.
By implication, the physical realisation of indefinite self-referential logical program such as the Halting problem relies on an external network of other external logical processors to prevent errors. 
We can no doubt build a system that executes the logic of an undecidable problem, but it is existence is guaranteed to be temporary by the ever rising tide of entropy.

The mathematical paradoxes uncovered by Turing and G\"{o}del do not bear any consequence on the physical world because they can never be truly realised in physics. 
I am uncertain whether we can bridge the Platonic divide, but I am certain that Universal Turing machines are not a viable avenue to do so. 
Like many other mathematical anomalies arising in the realm of forms, they are denied a physical existence by the laws of thermodynamics.

\bibliographystyle{unsrt}
\bibliography{FQXI_essay_2020}

\end{document}